# Current and Future Directions for Responsible Quantum Technologies: A ResQT Community Perspective

Adrian Schmidt[1*], Alexandre Artaud[2], Arsev Umur Aydinoglu[3], Astrid Bötticher[4,5], Rodrigo Araiza Bravo[6], Marilu Chiofalo[7,8], Rebecca Coates[9], Ilke Ercan[10], Alexei Grinbaum[11], Emily Haworth[12], Carolyn Ten Holter[13], Eline de Jong[14], Bart Karstens[15], Matthias C. Kettemann[4], Anna Knörr[16], Clarissa Ai Ling Lee[17], Fabienne Marco[18], Wenzel Mehnert[19,20], Josephine C. Meyer[21,22], Shantanu Sharma[23], Pieter Vermaas[24], Carrie Weidner[25], Barbara Wellmann[26], Mira L. Wolf-Bauwens[13,27], Zeki C. Seskir[1]

[1]*Institute for Technology Assessment and Systems Analysis, Karlsruhe Institute of Technology, Germany*
[2]*Department of Quantum Nanoscience, Kavli Institute of Nanoscience, Delft University of Technology, The Netherlands*
[3]*Science and Technology Policy Studies, Middle East Technical University, Turkey*
[4]*University of Innsbruck, Austria*
[5]*Friedrich Schiller University Jena, Germany*
[6]*Department of Physics, Harvard University, USA*
[7]*Department of Physics, University of Pisa, Italy*
[8]*National Institute for Nuclear Physics (INFN), Italy*
[9]*Commonwealth Scientific and Industrial Research Organisation (CSIRO), Australia*
[10]*Department of Microelectronics, Delft University of Technology, The Netherlands*
[11]*CEA-Saclay, France*
[12]*PushQuantum, Germany*
[13]*University of Oxford, United Kingdom*
[14]*Institute for Logic Language and Computation, Institute of Physics, QuSoft Research Center, University of Amsterdam, The Netherlands*
[15]*Rathenau Instituut, The Netherlands*
[16]*Quantum Science & Engineering, Harvard University, USA*
[17]*School of Business, Monash University Malaysia, Malaysia*
[18]*Technical University of Munich, Germany*
[19]*Austrian Institute of Technology, Austria*
[20]*Technical University of Berlin, Germany*
[21]*Dept. of Physics, University of Colorado Boulder, USA*
[22]*Dept. of Physics and Astronomy, George Mason University, USA*
[23]*Quantum Ecosystems Technology Council of India, India*
[24]*Ethics and Philosophy of Technology Section, Delft University of Technology, The Netherlands*
[25]*Quantum Engineering Technology Laboratories, H. H. Wills Physics Laboratory and Department of Electrical and Electronic Engineering, University of Bristol, United Kingdom*
[26]*Deloitte, Germany*
[27]*Geneva Science and Diplomacy Anticipator (GESDA), Switzerland*

Quantum technologies (QT) are advancing rapidly, promising advancements across a wide spectrum of applications but also raising significant ethical, societal, and geopolitical impacts, including dual-use capabilities, varying levels of access, and impending quantum divide(s). To address these, the Responsible Quantum Technologies (ResQT) community was established to share knowledge, perspectives, and best practices across various disciplines. Its mission is to ensure QT developments align with ethical principles, promote equity, and mitigate unintended consequences. Initial progress has been made, as scholars and policymakers increasingly recognize principles of responsible QT. However,

---

Corresponding author: adrian.schmidt2@kit.edu

more widespread dissemination is needed, and as QT matures, so must responsible QT. This paper provides a comprehensive overview of the ResQT community's current work and states necessary future directions. Drawing on historical lessons from artificial intelligence and nanotechnology, actions targeting the quantum divide(s) are addressed, including the implementation of responsible research and innovation, fostering wider stakeholder engagement, and sustainable development. These actions aim to build trust and engagement, facilitating the participatory and responsible development of QT. The ResQT community advocates that responsible QT should be an integral part of quantum development rather than an afterthought so that quantum technologies evolve toward a future that is technologically advanced and beneficial for all.

# 1 Introduction

Quantum technologies (QT) are evolving rapidly, promising groundbreaking advancements in computing, communication and sensing (Deutsch, 2020). However, these innovations come with significant ethical, social, and geopolitical implications that demand careful and timely consideration (Coenen et al., 2022; Gasser et al., 2024). The Responsible Quantum Technologies community (ResQT) was formed in 2021 as part of a workshop series to ensure that QT developments align with ethical principles, promote equity, and mitigate unintended consequences.

This paper systematically documents the current and future directions of the ResQT community. By examining the ethical, social, economic, geopolitical, and environmental impact of QT, this work provides an overview of ongoing efforts and emerging pathways for responsible QT innovation within the community. Specifically, it maps the current directions within the ResQT community, capturing the pressing questions across domains such as ethics, governance, sustainability, education, and geopolitics. Second, it identifies the envisioned future impact and outcomes of responsible QT efforts, including assessment strategies, ethical frameworks, and best practices in education and international collaboration. Lastly, it proposes the future priorities for achieving these goals to guide QT development as it becomes more integrated into societal infrastructures.

By synthesizing normative and empirical research, expert insights, and interdisciplinary collaboration of the respective authors, this work provides a reference point for scholars and policymakers for responsible QT innovation, ensuring that QT are formed with ethical foresight rather than reactive regulation. Through this work, the ResQT community seeks to create a foundation for future dialogue, action, and accountability in shaping the responsible development of quantum technologies.

# 2 Motivation and Context

The various QT have developed enormously in recent years: quantum computing hardware is developing rapidly (Wilhelm et al., 2025), the first quantum sensors are ready for the market (Kantsepolsky et al., 2023), and developments in quantum communication are visible (Schmaltz et al., 2025). The vast array of potential applications are attracting immense private and public investments, leading to international initiatives such as the Quantum Europe Strategy.[1] Other international efforts, such as the International Year of Quantum Science and Technology (IYQ)[2] are also emerging.

---

[1] https://digital-strategy.ec.europa.eu/en/library/quantum-europe-strategy
[2] https://quantum2025.org/

With QT starting to move out of the lab, researchers and policymakers are debating their real-world usage and impacts, including potential dual-use or military contexts (Krelina, 2025). Furthermore, there are significant global disparities in investment levels, which could lead to a widening gap in capabilities and competitiveness worldwide, the so-called *quantum divide* (Ten Holter et al., 2022). On a smaller level, within respective societies, certain population groups could be favored in the development of QT, despite the fact that learning materials are often freely available, what we may call the *local quantum divide* (Gercek & Seskir, 2025). These geopolitical, regional and local divides will potentially lead to further-reaching effects in diverse sectors and application areas. It is therefore essential to investigate the political, economical and societal implications of QT, so strategies for a responsible and beneficial operationalisation of QT can be developed (Gasser et al., 2024; Roberson, 2023).

Since there is already some evidence of negative consequences of QT, strategies need to be developed to anticipate and mitigate them. For example, a downstream impact is notable in terms of who gets access to the technology and for what reason, as well as the impact of decisions to invest in QT rather than other technologies or more immediate social needs. There is therefore a growing recognition and demand for adopting anticipatory governance approaches towards emerging technologies (OECD, 2024) due to lessons learned from social media and artificial intelligence (AI) governance (Quantum Flagship, 2025). In the history of technology assessment (TA), this situation is best identified by David Collingridge in his book *The Social Control of Technology* (1982), where he argues that early stages of technology development cycles are when the possibility of social control is the highest while the evidence of social impact is scarce or non-existent (Collingridge, 1982) a phenomena known as the "Collingridge dilemma." The Collingridge dilemma is a constitutive argument for the importance of research into potential social impact. It also highlights that research on these topics, the social impact, cultural disruptions, economic changes, changing power dynamics, etc. should evolve together with the technology, which we believe can best be practiced by keeping an active engagement with technical communities involved in developing these technologies.

As a response to these highly intertwined challenges and to foster the chances of positive outcomes resulting from a responsible and sustainable development of these technologies, the Responsible Quantum Technologies workshop series was launched, with an initial workshop organised by ZCS in 2021 at the Institute for Technology Assessment and Systems Analysis (ITAS) within the Karlsruhe Institute for Technology (KIT). The event became an annual event in Karlsruhe, with the fifth workshop held in April 2025. During these years, a community formed that was primarily focused on discussing the social, economic, geopolitical and environmental impacts of QT. This coalesced into a multidisciplinary environment consisting of researchers in STEM (mostly physicists and engineers involved in QT development), social sciences and humanities (participants from science and technology studies (STS), technology assessment (TA), philosophy, sociology, anthropology, political science and similar backgrounds), and education research (particularly physics education research) fields, as well as practitioners such as artists, corporate outreach and community managers, and equity diversity and inclusion (EDI) officers.

Through the years, the event series has hosted 138 speakers and panelists from 70 institutions with a wide range of backgrounds and gender-balance. Most participants were either professionals (46), professors (35) or doctoral researchers (31) and the number of participants is continuously growing (Figure 1). In terms of affiliation the Delft University of Technology (10), KIT-ITAS (7) and IBM (6) are the most prominent institutions with PushQuantum (5), Technical University of Munich (TUM) (5) and the University of Bristol (4) following. Although there are also participants from different world regions, reviewing the affiliations reveals a bias toward institutions from Europe, partly reflecting the workshop's location in Germany. This may indicate that Europe is currently investing more heavily in responsible QT initiatives, while other regions are comparatively less represented in this area. The community is aware of this imbalance, and resolving it is a focus for future activities. However, the affiliations do not reflect nationality or diversity of perspectives.

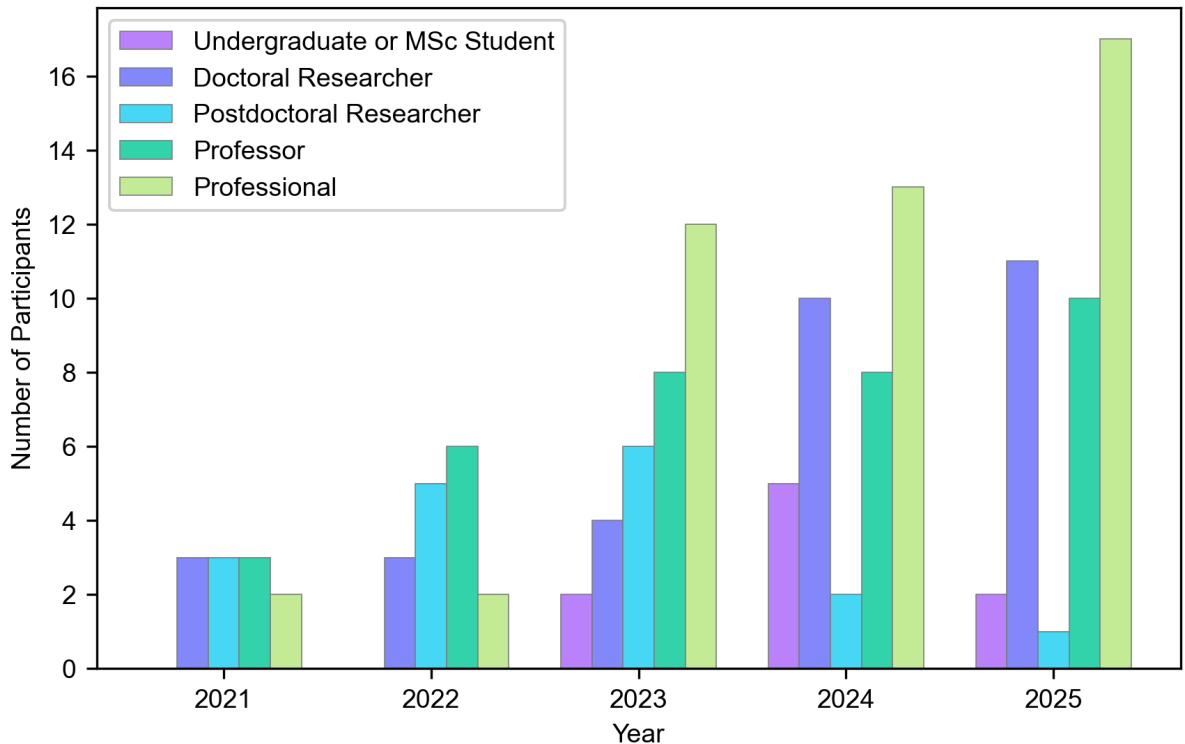

*Figure 1 - Demographics of participants of ResQT workshops*

ResQT is organised in a highly inclusive way. It has no rigid boundaries and only requires work on the impacts of QT. In particular, there is no aim to reach a common consensus on what responsible QT must be, which effects are most likely, or which focus is most important; instead, a plurality of opinions resulting in a multifaceted perspective on urgent problems and possible solutions are discussed openly.

Over these five years, a number of topics have emerged in which the ResQT community addresses the consequences of QT. These are not rigid but have developed iteratively from the experiences of the individual meetings and have been constantly reflected upon, consolidated, and expanded.

- In the context of **ethics** and **Responsible Research and Innovation (RRI)**, the fundamental ethical questions regarding the development, application, and use of QT are frequently discussed. These discussions include the learnings from other technology ethics, the implementation of RRI principles in the development process to address the challenges, benefits, and risks of QT development, in return generating societal trust and promoting the use of QT for the greater good.
- Within **equity, diversity, and inclusion (EDI)**, the reasons for inequality in QT are analyzed, and solutions to overcome these inequalities are researched. These include initiatives to widen access and reduce the quantum divide, increase representation and participation, and identify methods and initiatives to address these issues.

- Regarding **sustainability**, the environmental impact of current and future QT systems is being studied. This includes energy consumption, sustainable applications, and ways to implement sustainability measures in the development process.
- In the **geopolitical sphere**, the impact of international power dynamics on the development and access to QT is increasingly discussed. This includes analyzing current trends, exploring strategies to mitigate negative effects, and easing tensions through collaboration. At the same time, the dual-use nature of quantum technologies, spanning military and civilian applications, poses unique challenges for a 'responsible' quantum community, raising questions of democracy, power balances, and ownership of knowledge and technology.
- Within **education and workforce development** the most effective ways to learn and teach QT, including its societal and ethical dimensions, are being examined. Different methods for understanding QT are studied here, as well as how to train a global quantum workforce that includes people in underserved regions.
- In **outreach & science communication,** the questions of how to best reach and inform the general public about QT and how to include them in shaping its development are being discussed. Questions arise about which communication practices are accessible to a larger audience while remaining scientifically accurate and having a long-term impact.
- With **art-science interaction**, the tools to inform and educate people about QT are being expanded, as well as the tools to shape its development in an ethical and socially acceptable way. It's being discussed whether and how the arts can play a role in co-creating QT and how they might bridge gaps to societal groups that so far have not been included.
- Lastly, the question of how political decisions can be made to fulfill ResQT's goal of responsible, sustainable, and socially acceptable QT development is discussed in **governance**. This includes effective funding mechanisms, regulations, and international cooperation that serve QT development *and* society as a whole.

Nevertheless, the field and the community need to continue to evolve. The use cases of QT are becoming more concrete, the number of participants is increasing, and as these technologies are near readiness, the number of people potentially affected by QT development is increasing (Gercek & Seskir, 2025; Vermaas & Mans, 2024). It is therefore not surprising that the implications of QT and responsible development are being considered in a growing number of policy documents, including central strategies for developing and deploying QT adopted by the EU[3], the USA[4], the G7[5] and in a variety of hackathons[6,7]. This creates an opportunity for the community to pay more attention to other issues, such as the relationship between QT and other technologies such as AI, or the implications of recent geopolitical changes.

---

[3] https://digital-strategy.ec.europa.eu/en/library/quantum-europe-strategy
[4] https://www.quantum.gov/
[5] https://www.consilium.europa.eu/media/3l2f1jtr/quantum-en.pdf
[6] https://hackathon.nyuad.nyu.edu/
[7] https://qtric.sut.ac.th/quantathon2025/

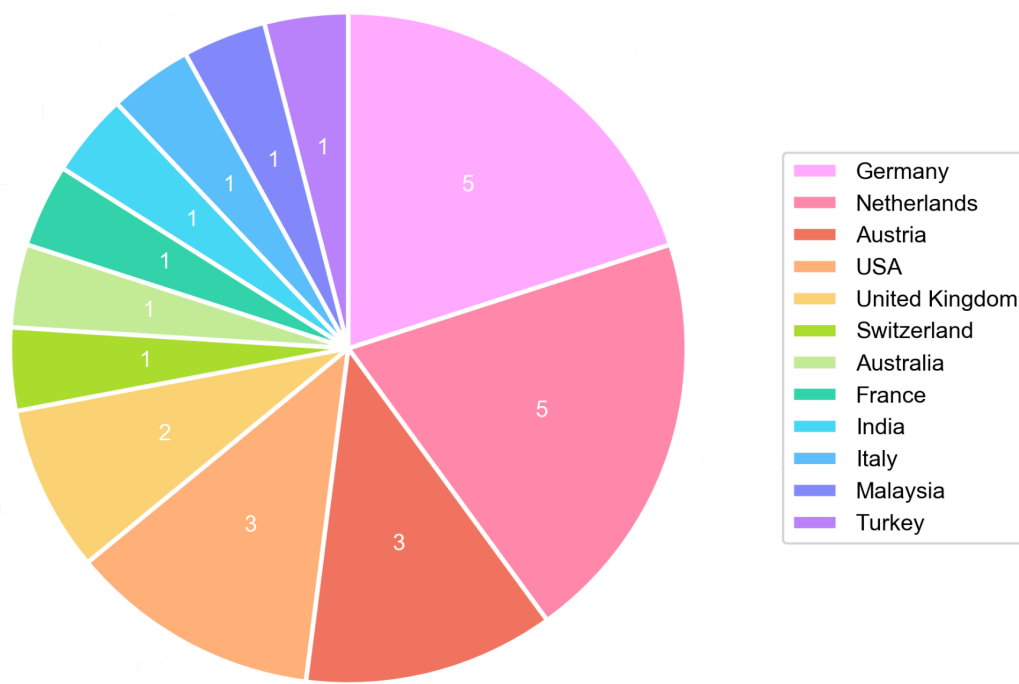

*Figure 2 - Affiliated countries of authors*

Members of the ResQT community are aware of such developments and continue to explore ways to address them. Until now, these discussions have largely taken place within workshops or events, without being published in a structured or accessible form. To ensure that both the ResQT community and related stakeholders, including researchers, policymakers, and societal actors, understand the importance of responsible research in QT, this knowledge must be collected, systematized, and shared (Umbrello et al., 2024). This effort requires three core dimensions which are the objectives of this community statement:

1) Mapping **current directions** within ResQT, to inform both internal and external audiences about ongoing research, current works, and existing expertise.
2) Articulating the **envisioned impact** of ResQT's work, outlining the societal, ethical, and scientific contributions it aims to make.
3) Formulating the **future priorities** that address currently underexplored or overlooked areas, ensuring that the field evolves in a responsible, inclusive, and forward-looking manner.

## 3 Methods

A qualitative approach was chosen to gather answers to the objectives. This means that only ResQT members working on different issues related to the effects of QT were involved in giving input for the objectives on their respective topics. The results were collected, summarised in thematic small groups, and published in the same structure.

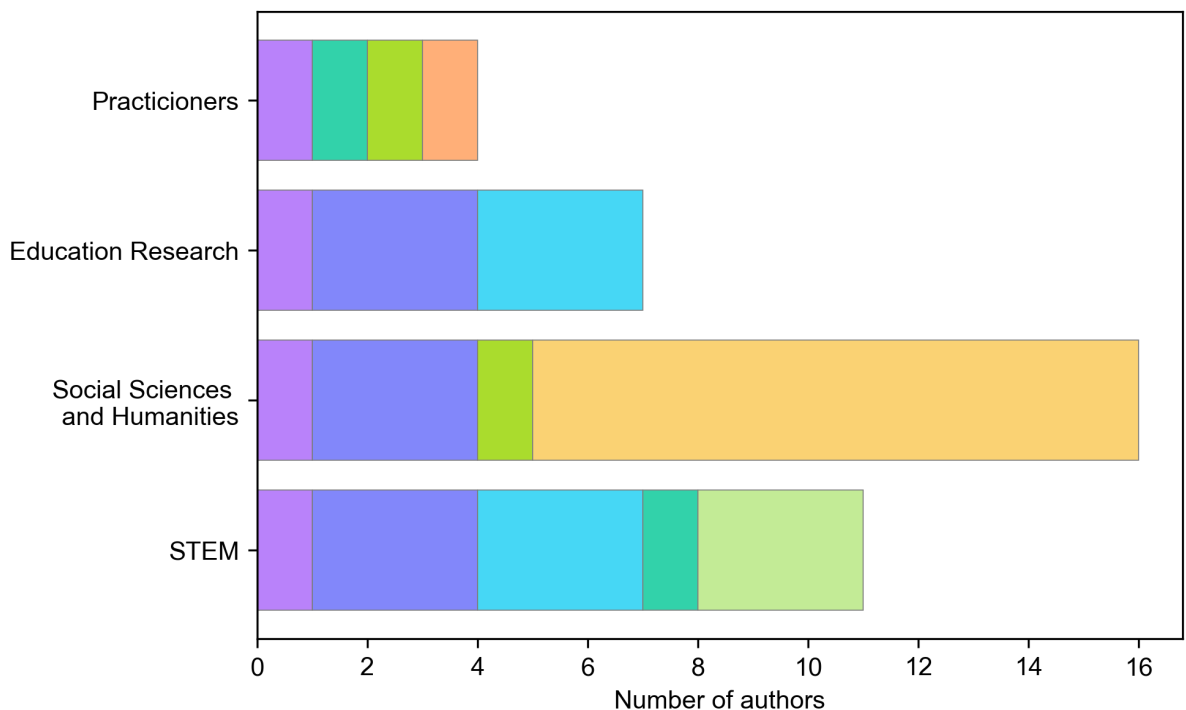

*Figure 3 - Distribution of authors across disciplines. Multiple selections per author are indicated by consistent colors across different disciplines.*

Specifically, all 138 participants of previous ResQT workshops were invited via email to provide input on the topics they were interested in through an online form. The topics were pre-defined by two of the authors on the basis of previous ResQT sessions: ethics, RRI, EDI, sustainability, geopolitics, education and workforce development, outreach and science communication, art-science interaction, and governance. Based on feedback from the authors, these were iteratively expanded to include sustainability, while ethics and RRI were combined due to the similarity of responses.

Twenty-five people participated and are listed as authors here. They are affiliated with 21 institutions and are based in 12 countries spanning four continents, as shown in Figure 2. Figure 3 shows that they come from diverse disciplines, including STEM, social sciences and humanities, education research, as well as practitioners like consultants, artists and community managers. Around half of the authors are affiliated with at least two disciplines. Therefore, the authors represent a good cross-section of the diverse ResQT community.

The authors provided, on average, input on two to three different topics. On the basis of further queries, nine people - AS, CALL, CW, EH, IE, MWB, PV, RC and ZCS - agreed to merge the contributions on the individual topics. Based on a predefined uniform structure, all those who had contributed to a topic were then included by the group heads in order to generate a corresponding text of about 600-800 words. These texts produced are the results presented. The final version of this work was reviewed and approved by all authors. As with the ResQT community itself, the purpose of this community statement was not to reach a final consensus, but rather to identify the general lines of current and future inquiry. Therefore, while all authors agreed on the main lines of action, they may have slightly different perspectives and personal objectives. Thus, a given author may not agree on all aspects included within the paper, and the paper contents do not necessarily reflect the opinions of the authors' various institutions.

# 4 Results

What follows are the collected contributions, which were curated and refined through feedback from all authors. The contributions are structured according to the three objectives to ensure coherence and focus. Although arranged in a specific order, as explained in Section 3, the topics are neither directly combined nor built upon sequentially, allowing each to stand on its own while leaving room for certain parallels. That is, multiple subsections discuss similar themes, and this has been left as-is instead of actively avoiding redundancy, in order to inform the reader and broader community of the overlaps between these topics.

## 4.1 Ethics and Responsible Research and Innovation

*Current Directions*

The ResQT community is actively addressing ethical questions of QT, with a particular need to focus on quantum computing, communications and sensing as dual-use technologies[8]. By drawing lessons from other fields, such as nanotechnology, classical computing, and AI, ethical challenges of QT applications are currently being identified and it is being explored to what extent ethical questions of QT are different (Shelley-Egan & De Jong, forthcoming). The work under *ethics and RRI* also includes the operationalization of ethics and RRI principles, identification of frameworks, tools, and methods best suited to address practical implications for researchers, entrepreneurs, and policy makers. Ongoing work also includes synthesizing those insights into a common understanding of what is meant by *quantum ethics*, including best practices in ethics and RRI of QT and the commonalities and distinctions between QT and other fields. Initial proposals for the definition of e.g. responsible quantum computing have been made (Mandelbaum et al., 2025).

Additionally, the ResQT community is exploring how to best engage with the general public and different stakeholder groups regarding QT, particularly with regards to demystifying the technology and understanding where resistance to its adoption may arise (McCrea et al., 2024; Ten Holter et al., 2023). This includes reaching out to a broader global ecosystem to counteract the tendencies of QT being perceived as strange and complex. The development of clear and transparent communication should help to foster public understanding of QT, in order to equip them to actively engage in discussions about challenges, risks and benefits. (de Jong, 2025). Such transparent communication is hindered by an over-reliance in the QT field on promises and hype that, by nature, do not promote honesty or lucidity, as well as the language describing quantum theory as, e.g., "impossible to understand" or "magic", which cannot be easily interpreted by the lay public (Grinbaum, 2017; Meinsma et al., 2023). Overall the current goal is that ongoing quantum innovation should align with societal values so QT can serve the public good responsibly and inclusively.

*Envisioned Impact*

In order for work in *quantum ethics* to be impactful, the ResQT community emphasizes a community-wide understanding and consensus of what *quantum ethics* is and why it matters. Where challenges arise, barriers to responsible QT should be highlighted and addressed. The outcomes the community desires focus on, first of all, a clearer understanding of what *quantum ethics* is, what key themes it (should) address and how it can be impactful.

---

[8] While this term has a broader meaning (Grinbaum & Adomaitis, 2024), in the quantum case it predominantly implies technologies with both civilian and military applications.

Desirable outputs of this research would be a ranking of the potential risks related to different applications of QT, guidelines for sustainable and open development of QT, and an evolving understanding of the societal impacts of QT. Additional desired outcome centres on practical guidelines for the operationalization of responsible innovation in QT, starting with awareness of potential risks, unintended consequences and mitigation strategies. For instance, this could include a rewriting of the six RRI dimensions (ethics, gender equality, open access, public engagement, science education, and governance (European Commission: Directorate-General for Research and Innovation, 2012) specific to the different aspects of QT, generating a checklist to be used when evaluating the impact of QT when making a decision, either at a scientific or political level.

*Future Priorities*

To obtain the impacts discussed above, the responses from the community can be broadly distilled into seven research questions to be addressed as future work: (1) What methods need to be developed for understanding the ethical challenges in QT? (2) Can we learn from past challenges and other emerging technologies? (3) What is best practice for ethics and RRI in quantum technologies? (4) How can it be best communicated, operationalized, and implemented? (5) How can the societal impact of QT best be understood and shaped towards what many refer to as "quantum for good"[9], i.e., the concept of creating positive societal, environmental, and economic impact through quantum technology? (6) How can we ensure that access is equitable and global? (7) What are the current and potential future opportunities and risks associated with QT, including its dual-use and military applications?

In order to be impactful, in answering these research questions, the ResQT community understands the need to move beyond abstract principles and values to their operationalization in the form of concrete design steps to be implemented by QT professionals. Thus, the future work in this area will focus on how to operationalize RRI and quantum ethics in the development and deployment of QT, including leveraging lessons from previous emerging technologies, while also ensuring the technology itself performs as expected. A priority in order to be impactful is to work closely with policy makers in order for ethics and RRI to be included into national strategies (Atladóttir et al., 2025).

## 4.2 Equity, Diversity, and Inclusion

*Current Directions*

The ResQT community notices that despite a rapid expansion of QT, disparities in access to resources, education, and opportunities - the quantum divide - remain a significant barrier to achieving equity, diversity, and inclusion (Wolbring, 2022). Based on insights from initiatives like the Strengthening and Entangling Global Quantum Roots (SEGQuRo) project and the Quantum Grassroots initiative[10], ResQT members are involved in addressing these challenges through grassroots community building, education, and ethical engagement.

Members of the ResQT community recognize the importance of gaining a detailed and context-sensitive understanding of the specific ways in which inequalities in quantum technologies arise, how they are sustained or exacerbated through systemic or structural mechanisms, and what strategies or interventions might effectively address them (Gercek & Seskir, 2025). For instance, what lessons from nanotechnologies or AI might guide inclusive QT development? How can QT participation be made equitable, considering global disparities in funding and representation? How might the increasing representation of women in the QT community influence EDI efforts (Genenz et al., 2025)?

---

[9] https://quantumforgood.eu/
[10] https://quantumdelta.nl/news/six-projects-funded-by-open-call-from-the-centre-for-quantum-and-society

Additionally, the ResQT community is investigating who currently has access to QT education and how it is perceived across diverse cultural and regional contexts. This includes reflecting on how educational content might better align with local knowledge systems, drawing inspiration from frameworks like ethnomathematics, which aim to decolonize STEM curricula by embedding cultural perspectives into technical education (Kabuye Batiibwe, 2024).

*Envisioned Impact*

A more inclusive and equitable quantum future should begin from the ground up. Grassroots initiatives play a key role by highlighting local barriers and developing community-specific solutions — but for lasting change, institutions need to take ownership and embed these efforts into their structures. This means supporting dedicated funding streams, such as scholarships or fellowships, that are explicitly tied to equity goals and open doors for underrepresented groups. It also means building frameworks that ensure EDI values are not just encouraged but formally integrated into how quantum technologies are governed.

In order to understand what works and where progress is needed, long-term studies of how equity and ethics evolve within quantum organizations should be carried out. These can help shape future action. Open-access resources designed specifically for underrepresented communities — along with hybrid events that remove travel or language barriers — can further reduce gaps in participation (Maldonado-Romo & Yeh, 2022). International collaborations that center ethics and shared responsibility will be key to shaping a quantum ecosystem that is truly global, inclusive, and fair.

*Future Priorities*

In the coming years, the ResQT community should continue expanding quantum education efforts to better serve marginalized communities and support cultural change across the field. A key priority is understanding where learners fall through the cracks in the educational pipeline and developing strategies to support them. Making quantum technologies more relevant to people's lived experiences—through applications in health, the environment, or local economies—can help engage communities often excluded from the mainstream QT discourse (Gercek & Seskir, 2025).

Another important objective is to address how QT priorities—often tied to military or high finance applications—may limit participation or reinforce existing inequalities. Research evaluation and funding structures should also be re-examined to avoid contributing to these disparities (Ten Holter et al., 2022).

Grassroots initiatives like SEGQuRo have shown how locally driven workshops, hackathons, and education projects can build meaningful QT engagement without exacerbating brain drain. These efforts identify culturally specific needs and create tools that are relevant and sustainable. Expanding such programs—especially in regions like Africa, South America, and Asia—will be essential to a globally inclusive QT ecosystem (Genenz et al., 2025).

## 4.3 Sustainability

*Current Directions*

Quantum computers are primarily being designed to surpass classical systems in computational power for niche applications. Recently, a growing research focus explores their potential to provide an energy advantage (Jaschke & Montangero, 2023). Establishing a standard for benchmarking "performance per Watt" is vital for meaningful assessments. The Quantum Energy Initiative (QEI)[11] and its associated IEEE working group[12], in which members of the ResQT community are involved, are centralising the effort to both improve physical resource efficiency (incorporating reduced non-renewable material cost and waste) and create an energy standard (Auffèves, 2022). These initiatives currently focus on creating these measures and researching how they could be implemented broadly across the field.

Gathering input from high-performance computing centers - anticipated to be primary users of quantum computers and already operating under initiatives to optimise energy efficiency - would provide valuable perspectives on, e.g., the trade-offs between wet and dry cryogenics, weighing energy consumption against operational simplicity, and support in conducting larger lifecycle analyses of quantum technologies (Cordier et al., 2025). Beyond technical research, the ResQT community looks to investigate social strategies, not strict restrictions, to keep sustainability forefront, for example, by trialling measures such as artificially expensive energy with internal reinvestment, fostering friendly lab competitions, or reporting $CO_2$ equivalent costs in publications[13].

*Envisioned Impact*

Any technical success in achieving energy efficiency must be coupled with safeguards to ensure its benefit is actually felt and it doesn't instead suffer from induced demand or rebound effects (Karakaya et al., 2024). Transparency in reporting the environmental and energy costs of quantum applications versus their benefits is critical for determining whether the technology delivers net sustainability gains. At this pivotal point of history, it is important to be part of the solution, instead of adding to the problem.

On the application and adoption side, quantum has been touted as a promising tool to solve relevant issues such as energy grid optimization (Ajagekar & You, 2019). The question remains, however: what guarantee is there that quantum solutions would actually be implemented where others have not been? The desired future impact here would be the development of effective policies, improved communication between stakeholders (including the public), realistic analysis, and concrete roadmaps that provide clarity on the practical applications of QT and their way into the market.

Achieving these goals will require sustained efforts and collaboration. Every stakeholder must be transparent and include some consideration of sustainability in their work. Although all deep tech is inherently high-risk, the field can at least act as a blueprint for how technology should be responsibly developed. Proactive action now, while solutions are still malleable, will help avoid resource bottlenecks and ensure the longevity of QT.

---

[11] https://quantum-energy-initiative.org/
[12] https://sagroups.ieee.org/qei/
[13] https://scientific-conduct.github.io/

*Future Priorities*

Many facets to the issue of sustainable development are not unique to quantum, and existing regulations should be adapted to include quantum technologies. Ensuring sustainability can have additional benefits for the field (Root, 2025). Supply chain transparency aids in addressing scalability challenges. Carbon pricing - already relevant to classical cloud computing - must complement claims of quantum energy advantages in order for any realization to be economically viable.

As quantum research moves from theoretical models to impactful real-world applications, the ResQT community should continuously assess how stakeholders can be incentivised to prioritise projects with tangible, long-term benefits over short-term financial returns. Integrating innovations with legacy infrastructures and aligning them with political will is crucial. Frameworks[14] and incentives should be explored to guide companies toward sustainable decision-making and promoting research into sustainable use cases. For instance, portfolio optimisation holds potential ESG (environmental, social, governance) benefits, but only a fraction of research in this area incorporates ESG considerations, underlining the need for more comprehensive sustainability integration in early quantum use cases.

Additionally, the ResQT community believes more attention on the intersection of quantum workforce development and sustainability is merited. What role can workforce policies play in fostering inclusive job markets, through which a wider range of sustainability challenges are addressed (Smith-Doerr et al., 2017), particularly those related to human impacts within the supply chain? How can economic benefits from quantum advancements contribute to broader societal sustainability, including public support for climate policies and environmental justice?

While numerous challenges arise during the growth stage, sustainability should not be viewed as an additional burden. Instead, it should be embraced as a guiding principle - a green blueprint for conducting business. Sustainability is a strategic approach rather than a problem to be solved. Therefore, it is crucial to integrate sustainable practices gradually and thoughtfully.

## 4.4 Geopolitics

*Current Directions*

Research on the geopolitics of QT, especially within the ResQT community, can be grouped into analytical, normative, and methodological strands. Analytical work investigates how QT is framed as a sovereignty technology and a national security concern, exploring whether this framing acts as a self-fulfilling prophecy that fuels nationalism and global competition (Coenen et al., 2022). It considers how geopolitical shifts impact the organization of the QT ecosystem and the implications for countries and communities not yet engaged in QT development, particularly in the Global South (AWO, 2024). There is also concern about whether QT may be entangled in future geopolitical rivalries, with the US-China dynamic often used as a focal case. This includes scenario planning for states, considering access to knowledge, workforce, materials, and funding.

Normative research focuses on identifying ways to mitigate the negative consequences of these geopolitical dynamics (Possati & Vermaas, 2025). It asks how tensions can be mediated, how openness in research impacts outcomes, and what role international collaboration should play. Methodological reflection deals with how to approach global technology assessment and scenario planning in this complex landscape and how to integrate these into proposals for political governance of QT that are inclusive, forward-looking, and globally relevant (Possati, 2024).

---

[14] https://www.pushquantum.tech/sustainability-guide

*Envisioned Impact*

The aim of the community's research is to foster anticipatory international, multilateral dialogue, increase openness, and promote collective awareness about the geopolitical aspects of QT. It should contribute to a well-articulated discourse that resonates within both policy and quantum communities. Making the potential geopolitical risks of QT applications explicit and ranking them can inform practical strategies and interventions. In collaboration with governments, this research should support the development of normative responses to geopolitical tensions and propose actionable guidelines.

The overarching goal is to ensure that QT does not become a tool for division or conflict but instead offers opportunities for cooperative advancement. By articulating the risks and initiating informed public and policy debates, research in this area aims to prevent the onset of a new cold war driven by technological rivalry and mistrust. Research may also focus on effective means to this prevention, such as an international treaty on the non-weaponization of QT aimed at regulating its development responsibly. Through clearer communication and stronger alliances, QT's future can be guided by shared values rather than hardened geopolitical boundaries.

*Future Priorities*

To achieve this vision, future research must address several strategic questions. One focus is on how technical standards in QT could contribute to the geopolitical division of the world and whether these standards can instead promote global cooperation. Another line of inquiry examines the extent to which QT research is siloed or shared and how to ensure that access is equitable and independent from political agendas. A central concern is how QT can be made available to the majority of the global population, emphasizing inclusivity over exclusivity (Vermaas & Mans, 2024).

Research must also explore how policy frameworks can keep pace with the rapid technical developments in QT, remaining agile and effective, and support policy efforts towards the non-weaponization of QT. These questions are integral to shaping a QT ecosystem that is ethically grounded, globally inclusive, and resilient to escalating geopolitical pressure.

### 4.5 Education and Workforce Development

*Current Directions*

Given the disruptive potential of QT, a wide range of societal stakeholders–including policymakers, industry leaders, educators, and the public—requires varying levels of understanding to make informed decisions, foster innovation, and anticipate societal impacts. As a near-term starting point, the community is generally interested in understanding how students reason and learn about QT, as well as the development of research-based curricular materials and assessment tools. There is a robust body of literature on how students reason about quantum mechanics (Borish & Lewandowski, 2024; Modir et al., 2017; Singh & Marshman, 2015). It remains to be seen, via discipline-based education research (DBER) (Schweingruber et al., 2012), how much of this work carries over to QT, including experimental work in QT (Borish & Lewandowski, 2023) and teaching QT to non-physics audiences (Coecke & Kissinger, 2017; Seskir et al., 2024). Likewise equitable and reliable tools to measure the understanding of QT are currently being developed, like the Quantum Computing Conceptual Survey (QCCS) (Sadaghiani & Pollock, 2015); more such tools will be needed in the future as the field develops. Equally important, the community is developing assessment tools that measure not only conceptual understanding but also how well students are equipped to navigate the societal and ethical dimensions of QT (Meyer et al., 2022).

Additionally, the community is investigating the effect of metaphors for developing quantum intuition and understanding, particularly with regards to their scientific accuracy (Meinsma et al., 2025). Similarly, we wish to understand the effects of quantum hype on education (Meyer et al., 2023), both positive and deleterious, especially in the context of ethics and equity of access. This includes incorporating ethics and responsible innovation into QT education, but also broadening the reach of different forms of quantum education into communities that are traditionally underserved (Genenz et al., 2025). That is, we must widen the global reach of QT education to overcome the rising quantum divide both within and across nations (Gercek & Seskir, 2025; Ten Holter et al., 2022). Finally, the community is focusing on how to develop and retain a quantum-literate workforce (Aiello et al., 2021; Greinert et al., 2024; Kaur & Venegas-Gomez, 2022). This includes understanding how quantum education competes with other forms of technical education, what kind of skills are needed for a quantum literate workforce, and whether there are certain required skills that might differ between different QTs.

*Envisioned Impact*

Broadly, the community wants to see, as concrete outcomes, a robust core of literature on learning and cognition for QT students (Donhauser et al., 2024), emphasizing teaching methods and learning metrics that incorporate ethics and responsible innovation within them. Based on lessons learned from other fields (e.g., neuroscience, AI), research-based curricula emphasizing *quantum ethics* and social responsibility must be developed. These in turn could serve as a blueprint for other technologies. The community wants to see these quantum education programs developed in an inclusive, accessible, and equitable way, including input from underserved communities (Arrow et al., 2023; Rosenberg et al., 2024). This means involving scholars from communities in the Global South, but also actors (e.g., social workers) in regions who are leading QT development. These programs may not be 100% scalable, but where possible, lessons learned and scaling methods should be shared among the community. Finally, there is a need to track the progress of QT-interested people through their educational and career pathways, which will inform future developments in QT education and workforce training.

*Future Priorities*

In the longer-term, all aspects listed above persist, showing that these cannot truly be solved in the near-term. This work must be done in a responsible and ethical manner, with an emphasis on learning from other disruptive technologies (e.g., AI). There is also an urgent need to diversify the DBER research pool so that research institutions in the Global South are more involved, thus meeting the needs of a future global quantum workforce (Kanim & Cid, 2020).

The QT community must also consider education beyond (quantum) scientists and engineers. That is, the community will need to reach a broader audience and understand how these people can contribute to QT, especially when considering ethics and responsible innovation. There is a need to educate people both in other fields and within the general public, both formally and informally. This could serve to lower the perceived barrier to entry, but we must research other means to lower this barrier.

Finally, the community is interested in scaling up educational programs in an equitable way, including understanding who should bear the cost burden. This calls for work in understanding the efficacy and applicability of alternative paradigms for understanding quantum mechanics (Seskir et al., 2022), including Quantum in Pictures (Coecke & Kissinger, 2017) and Explorable Explanations[15].

### 4.6 Outreach and Science Communication

*Current Directions*

The ResQT community notices that there is still a missing consensus regarding what is considered optimal outreach and science communication in QT (Chiofalo et al., 2024; Gutorov et al., 2025; Meinsma et al., 2024). This includes explicit content as well as the level of detail to address, depending on the specific audience. Therefore, the community is currently exploring which practices work and which do not. This includes understanding the misconceptions of different groups to effectively explain opportunities, risks and ethical implications of QT (Seskir et al., 2024). A major concern is how communication failures can create space for misleading or exaggerated claims to thrive, distorting public understanding and expectations. In particular, the need to achieve the necessary technical depth while using inclusive language is a current key research aspect.

Metaphors can play a crucial role in science communication of QT (Archer, 2022; Godoy-Descazeaux et al., 2023; Grinbaum, 2017). The community is therefore pushing beyond the traditional picture of Schrödinger's cat. The goal is to find more meaningful metaphors and analogies that convey quantum principles without oversimplifying them.

Additionally, mechanisms to support sustainable outreach and science communication efforts are being developed. This includes developing strategies that empower societal stakeholders like actors from the arts, museums, the media, or local communities in participating in the discourse. By including different views and designing tools or metaphors together, science communicators and outreach practitioners can have ideas at hand to reach the public effectively (Genenz et al., 2025; Seskir et al., 2024).

*Envisioned Impact*

One of the community's central goals is to enable public understanding of QT through transparent and inclusive communication practices. By striking the right balance between hype and skepticism, the community aims to equip society with the knowledge necessary to discern realistic promises from overblown claims (Roberson, 2023). This approach could ensure that QT's potential is appreciated while also fostering critical engagement with its risks and limitations (Skelton et al., 2025). Collaborative platforms could track evolving narratives while amplifying diverse voices, including contributions from students and stakeholders in underrepresented regions, ensuring a holistic understanding of QT.

A key future goal is the creation of innovative outreach formats that surpass traditional methods like museum displays or school lessons. These formats could foster inclusive engagement, integrating gender-sensitive approaches to inspire future generations of scientists and ensuring diverse societal participation. Workshops on science communication would become a standard practice, equipping researchers and different kinds of stakeholders to communicate QT's complexity clearly and accessibly. The desired outcome is a global community that understands QT's societal implications, supports responsible development, and actively participates in addressing risks, opportunities, and ethical challenges, ultimately guiding QT towards equitable and sustainable innovation.

---

[15] http://explorabl.es/

*Future Priorities*

To achieve these objectives, ResQT recognizes the need to broaden and analyze the current discourse on QT while developing long-term visions for effective outreach and science communication. Central questions are how narratives in the QT community, as well as public perceptions, shape QT development and influence governance, funding, and societal engagement (Meinsma et al., 2023; van de Merbel et al., 2024). Researchers are keen to understand how public narratives have historically shaped technology governance and how specific features of QT might alter these dynamics. Particular attention must be paid to how inadequate or misleading science communication contributes to the spread of pseudoscientific claims in the field of QT, potentially distorting public expectations and policy decisions. Additionally, the question arises as to whether reasoning based on quantum principles differs from other types of reasoning and whether the wider adoption of QTs might change the way humans understand the world in the future.

Expanding the thematic scope of QT discourse is essential. The focus should extend beyond computing to include applications in communication, sensing, and other domains as well as the integration of QT into other domains as high performance computing. This includes addressing ethical concerns like dual-use scenarios in military contexts and ensuring global equity in accessing QT as it becomes increasingly costly. Here, the questions should be investigated as to what access to QT can mean, who the actual users are going to be, if there are meaningful use-cases for all individuals, and which use-cases will affect which stakeholder groups. These efforts should aim to engage a broader audience, including stakeholders who might have been previously excluded from QT dialogues.

Reaching these diverse audiences requires a variety of communication strategies. Therefore, outreach must be tailored to specific groups, such as policymakers, educators, and the public. Strategic communication should also address misleading or exaggerated claims about QT by providing clear, accurate, and accessible information. By addressing diverse perspectives, ResQT seeks to foster inclusive and informed discussions, aligning technological development with societal needs and expectations.

## 4.7 Art-Science Interaction

*Current Directions*

Current research in the ResQT community also focuses on the emerging field of art-science interactions in quantum technology (QT). This includes tendencies toward "quantum art"[16] and its potential influence on public perception and QT development (Caro & Murphy, 2002; Crippa et al., 2024; Putz & Svozil, 2017). Researchers are working toward understanding what quantum art is, identifying key practitioners, and analyzing its evolution[17]. They are also contributing to the growing body of theater plays that explicitly or implicitly address quantum science (Skelton et al., 2025). They also explore how art can communicate QT's complexities to diverse audiences while preserving scientific integrity and advancing new forms of understanding (Cattan et al., 2024; Gaunkar et al., 2024; Heaney, 2019).

---
[16] https://medium.com/qiskit/theres-a-burgeoning-quantum-art-scene-76119cca7144
[17] https://medium.com/@adrian.schmidt.kit/quantum-computers-and-art-how-and-why-20d9c979be13

One central focus involves the study of QT representations in popular culture, especially science fiction (Possati, 2024). This could mean building narratives to authentically reflect the development of QT (Suter et al., 2024). It might mean investigating the role of metaphors, especially when it comes to how new metaphors could be developed to provide a deeper understanding of quantum phenomena (Meinsma et al., 2023).

ResQT researchers are looking beyond aesthetics by investigating the capacity of quantum art to engender an immersive experience that exceeds the epistemic status quo and present constraints in engagement (Thomas, 2018). The community is fostering and facilitating collaborations between artists, scientists, and stakeholders, as well as exploring how quantum art could shape the future of QT[18] (Ferreira et al., 2025). This includes stimulating innovation, inspiring the imagination, and providing a lens on ethical reflection into quantum technologies.

*Envisioned Impact*

The ResQT community envisions a future where art is integral to the engagement of diverse communities in the development of QT (Voss-Andreae, 2010). Researchers could leverage the arts to reach audiences traditional scientific communication could not reach while empowering them as knowledge co-creators. Art can provide more than a single way of accessing the technical intricacies of QT to individuals with varying thought styles, fostering a broader range of understanding and engagement.

At the end of the day, bringing together QT and art is not just about science communication[19]. One key objective would be to establish collaborative projects for artists to work on directly with societal groups wherein art might be the medium for involving these communities in the co-creation of QT. Such a participatory approach democratizes the development process, by ensuring that a broad spectrum of societal perspectives could inform QT's evolution. Developing best practices for integrating art into QT research and outreach will underline the transformative potential of such collaborations.

Ultimately, the ResQT community aspires to broaden QT's opportunities beyond obviously reproducing what the technical community already knows or understands about the possibilities of QTs[20]. By embedding art-science partnerships into QT development, the community aims to create new narratives, amplify the societal impact of quantum innovation, and align it with diverse cultural, ethical, and human values, to foster a more inclusive and responsible approach to quantum advancement.

*Future Priorities*

The ResQT community should delve deeper into the intersection of art and science to explore how these collaborations shape both expert and lay understanding of QT. The first critical area of focus is on determining how art can make QT more visible and accessible without reducing, oversimplifying or mystifying. The deployment of certain approaches, such as gamification (Piispanen et al., 2025; Seskir et al., 2022), could motivate the collaboration between STEM and non-STEM experts.

---

[18] https://www.goethe.de/prj/lqs/en/index.html
[19] https://arts.cern/exhibition/quantum-visions/
[20] https://medium.com/original-philosophy/if-quantum-physics-is-queer-what-does-it-mean-for-quantum-technologies-9e1ca6c2f674

The second critical area of focus concerns how artists could be effectively included in the co-creation of QT as equals. Artists should be able to bring their artistic practices into facilitating non-deterministic developmental trajectories in QT, thereby reconstituting cultural narratives around QT. Such contributions might foster the identification of methods and frameworks for the meaningful integration of scientific innovation and artistic expression.

Therefore, it is crucial for the ResQT community to consider how art-science activities could influence the eventual progression of QT while also shaping public perception. By constructing visions and goals that do not merely reproduce standard narratives, the community can converge into a more nuanced comprehension of QT's potential. Non-deterministic trajectories and cultural narratives are outcomes and driving forces of art-science interactions, thereby shaping decisions necessary to responsible QT development.

### 4.8 Governance

*Current Directions*

The governance of QT is challenging for various reasons, including the uncertainty within the technological development trajectory, the speed of development, and the tension between too-early vs. too late regulatory efforts. The ResQT is currently addressing a number of governance-related questions, which are largely related to the topics above.

One major focus is on building effective funding mechanisms that not only promote the development of QT themselves, but also support the responsible development and application of QT. This includes designing strategic and detailed funding mechanisms that effectively integrate the needs of different stakeholders. A key question is how far civil society actors can be actively involved, alongside the integration of companies and researchers.

Aligned with the vision of *Quantum for Good*, as suggested and promoted by the UN and the OECD, the community critically reflects on whether regulating specific QT developments will safeguard progress or, conversely, risk constraining the long-term evolution of the field. (Perrier, 2022). In this context, the levels of government action need to be taken into account to both reduce the risks of QT and exploit any available opportunities. Concrete efforts, situated both within and beyond the community include the ELSPI framework (Kop, 2023) suggesting the utility of simultaneously considering safeguarding, engaging, and advancing while approaching these phases of concrete developments simultaneously in different ways. Key considerations also include the further development of anticipatory governance mechanisms (Abboy et al. 2025), appearing to be more efficient than early stage regulation, enabling the community to adapt measurements for the responsible and sustainable development early on.

To examine how to align QT developments with societal values, the community concentrates on researching the influence of global governance structures (Atladóttir et al., 2025), the role of international cooperation within the dispersion of knowledge as well as global technology assessments and ecosystem analysis of QT (Parker et al., 2022; Possati, 2024).

*Envisioned Impact*

The themes that emerged within the governance context align with many of the topics discussed in the other sections of this paper. One such theme is bridging the quantum divide - both within Europe and on the global scale - to ensure greater collaboration and idea sharing and speed up QT development (Gercek & Seskir, 2025). An important objective is to include value-based aspects in the innovation system, shifting the focus from industrialization to societal goals.

Given the dual-use capabilities and significant known risks of QT, comprehensive risk assessments of QT are essential. These should include a tool to detect, analyze, and address future risks, without exaggerating unnecessary worries.

Key aspects of responsible QT development are still often overlooked - among them sustainability, which is central to ensuring long-term viability. Establishing sustainability guidelines would support developers, decision-makers, and producers of quantum applications in giving appropriate weight to environmental considerations. At the same time, principles of openness, shared best practices, and transparency can foster fairness and trust, creating a more balanced foundation for the field as it grows.

Finally, strengthening the talent pipeline can serve as a governance mechanism to connect education, research, and practice. Providing materials and resources that keep citizens informed about QT developments and responsible career pathways not only prepares the next generation of researchers but also helps align community values and practices.[21].

*Future Priorities*

To advance ResQT's mission, particular emphasis must be placed on bridging the quantum divide, developing resilient governance frameworks, and designing effective funding structures. Equally important are critical examinations of academic colonialism and sustained efforts to raise awareness among policymakers, as these dimensions are pivotal for ensuring an equitable and globally responsible quantum future.

Regarding the quantum divide, more attention must be given to the power dynamics of dominant companies and governments and on the impact of their private access to QT platforms. An important research objective is how to best inform the QT community about the opportunities of QT while effectively reducing the entry barriers. This issue is partially connected to academic colonialism. Although difficult to navigate, research should explore how support from the Global North can be reframed—not as aid to those "in need," but as an equitable collaboration. One approach might be a better exchange between academic institutions on how to collaborate regionally and globally for developing a strong QT field worldwide.

Resilient governance mechanisms are essential to address the dual-use capabilities of QT and the profound ethical, legal, and societal challenges they pose. While quantum encryption may provide unprecedented security for financial and medical data, the same capability could also destabilize international relations by rendering existing cybersecurity infrastructures obsolete. Meeting these challenges demands an integration of technical analysis with geopolitical dynamics and intergovernmental cooperation, e.g., by methods from technology assessment, to ensure the responsible advancement of QT.

---

[21]First examples of such work in a pioneering phase include https://www.quantiki.org/ and www.quantworld.org

Lastly, to further develop funding strategies such as new "first of a kind" and "public interest" funding strategies alongside rising awareness for the strategic impact of QT are central to the effective employment of the community efforts. Balancing (responsible) QT funding and other high-cost societal beneficial projects requires thoughtful discussion. Best practices should be developed for informing policy makers about potential impact of QT and how to strategically fund the analysis and mitigation of future QT risks. Questions may also arise about whether existing governance frameworks or forms of "real-time governance" could lead to informed decisions, so further developments in QT can be safeguarded.

# 5 Conclusion

Responsible quantum technologies are at a crossroads. This paper synthesizes insights from the ResQT community, reflecting current research, diverse perspectives, and emerging priorities. While this paper offers a valuable overview of the topics and pathways discussed in recent years, it should be understood as a snapshot shaped by the perspectives of its contributing authors - yet one that can be seen as a significant representative of the community. The eight topics outlined here offer guidance not only for the community itself but also for adjacent researchers, practitioners, and policymakers working at the edges of the field.

Several concerns cut across these topics. The community emphasizes the risk of a growing quantum divide - disparities in access to QT education, funding, and applications - and calls for equity, diversity, and inclusion to remain central. Sustainability is another pressing issue: both technical innovations to reduce environmental impact and policies that prevent QT from deepening existing inequalities are required. Trust and engagement are equally vital, ensuring that QT develops in ways that benefit all. Lessons can and should be drawn from past technologies such as AI and nanotechnology, where governance struggled to keep pace with innovation.

While this work underscores the necessity and potential of ResQT, it reflects the nascent stage of QT. This paper, therefore, is an intermediate step in recognizing the accomplishments so far as well as the future imperatives. The next step is to move from identifying problems to implementing responsible innovations in practice. That means fostering inclusive dialogue among all stakeholders, creating collaborative initiatives, and developing a truly multidisciplinary approach. Moreover, while the substantial number of authors effectively represents the currently active ResQT community, the number of researchers is continuously rising, and certain perspectives - for example, from non-European countries - remain to be included more directly.

Looking ahead, the rapid evolution of quantum technologies makes it imperative that responsibility is not treated as an afterthought but as a guiding principle. This paper has outlined the ethical, social, economic, geopolitical, and environmental dimensions of QT, highlighting both current directions and future priorities identified within the ResQT community. By bringing together normative and empirical research, expert insights, and interdisciplinary collaboration, it offers a reference point for scholars, practitioners, and policymakers committed to ensuring that QT development aligns with ethical principles, promotes equity, and mitigates unintended consequences. In this way, the ResQT community aims to establish a foundation for sustained dialogue, collective action, and accountable innovation as quantum technologies become increasingly embedded in societal infrastructures.